# The latent structure of global scientific development

Lili Miao[1], Dakota Murray[1], Woo-Sung Jung[2,3], Vincent Larivière[4,5,6], Cassidy R. Sugimoto[7], Yong-Yeol Ahn[*] [1,8,9]

[1]*Center for Complex Networks and Systems Research, Luddy School of Informatics, Computing, and Engineering, Indiana University Bloomington, IN 47408, USA.*

[2]*Department of Industrial and Management Engineering, Pohang University of Science and Technology, Pohang 37673, Republic of Korea.*

[3]*Department of Physics, Pohang University of Science and Technology, Pohang 37673, Republic of Korea.*

[4]*École de bibliothéconomie et des sciences de l'information, Université de Montréal, Montréal, Québec H2L 2C4, Canada.*

[5]*Observatoire des sciences et des technologies, Université du Québec à Montréal, Montréal, Québec H2L 2C4, Canada.*

[6]*Department of Science and Innovation-National Research Foundation Centre of Excellence in Scientometrics and Science, Technology and Innovation Policy, Stellenbosch University, Stellenbosch 7600, South Africa.*

[7]*Luddy School of Informatics, Computing and Engineering, Indiana University Bloomington, IN 47408, USA.*

[8]*Network Science Institute, Indiana University, Bloomington (IUNI), IN 47408, USA.*

[9]*Connection Science, Massachusetts Institute of Technology, Cambridge, MA 02139, USA.*

**\*Email:** yyahn@iu.edu

## Abstract

Science is essential to innovation and economic prosperity. Although studies have shown that national scientific development is affected by geographic, historic, and economic factors, it remains unclear whether there are universal structures and trajectories of national scientific development that can inform forecasting and policymaking. Here, by examining countries' *scientific 'exports'*—publications that are indexed in international databases—we reveal a three-cluster structure in the relatedness network of disciplines that underpin national scientific development and the organization of global science. Tracing the evolution of national research portfolios reveals that while nations are proceeding to more diverse research profiles individually, scientific production is increasingly specialized in global science over the past decades. By uncovering the underlying structure of scientific development and connecting it with economic development, our results may offer a new perspective on the evolution of global science.

**Introduction**

Science has experienced rapid transformation due to increasing scientific capacity across countries. During the Cold War, the USSR and the United States competed in science; the collapse of the USSR in the late 1990s and the concurrent rise of China on the international stage significantly altered power dynamics in science. Whereas China only accounted for 5% of scientific publications in international indexes in 2000, it became the most productive country in the world by 2018, surpassing US scientific production (1–3). The increase in scientific capacity was also coupled with Asia's economic acceleration: for example, the rapid expansion and intense industrialization of the "Four Asian Tigers"— Hong Kong, Singapore, South Korea, and Taiwan—occurred during this time (4, 5). These rapid transformations provide an opportunity to examine the relationship between economic and scientific development and to test theories of universality in this relationship.

Studies have examined how the interplay between geography (6, 7), history (8), existing scientific strengths (9–12), and economic conditions (13–15) influence scientific development. Chile, for example, exemplifies the influence of geographical opportunities on national knowledge production: despite relatively low scientific investments (16), Chile's unique mountainous and remote terrain made it ideal for astronomical observatories, a comparative advantage that allowed the nation to become an international hub in the field (17, 18). By contrast, South Korea, with its heavy investment in science and technology (19, 20), has experienced diversified scientific expansion, developing into a science and innovation powerhouse (4). Institutional organization and investment are also potential factors. For example, May (21) compared research organizations in France and Germany—where research institutes such as the CNRS and Max-Planck play have a central role—to the United States and the United Kingdom—which centralizes basic research in universities and engages students—and concluded that the former structure negatively affects research activity.

In contrast to localized explanations of national scientific development, several scholars have attempted to develop universal frameworks. For example, Comte argued that science develops along a natural trajectory from high-consensus physical sciences towards more complex, low-consensus social sciences (22). Basalla took a colonial perspective, arguing that scientific development of non-Western countries generally undergoes three phases: countries first provide resources for Western scientists, then transition to a replication model—in which science develops following the institutions and traditions from scientifically established nations—and culminate with scientific independence, often obtained with mixed success (23). In Basalla's model, phases of development also affect the research specialization of nations. For instance, in the first phase, disciplines are descriptive in nature and strongly tied to natural resources and exploration, the second phase has a stronger focus on experimental domains. Despite acknowledging that his model was a "heuristic device" and that the environment in which research is conducted should be taken into account (24, 25), Basalla's model remained criticized for being Eurocentric and insensitive to cultural factors (25).

Despite criticisms, some alignment between these theories and the observed phases of development has been observed. Moya-Anegón and Herrero-Solana (26) classified countries into three groups based on their research specialization and showed that high GDP countries specialize in biomedicine, formerly communist countries specialize in basic science and engineering, and less developed countries specialize in agriculture. Cimini et al. (27) arrived at a different conclusion, showing that—rather than specializing—technologically-leading countries have been active in a diversity of scientific domains.

Economic complexity framework provides a useful lens to evaluate economic and scientific transformations from a global perspective (28). This approach employs quantitative methods to predict and explain economic trajectories of geographic regions (29–31), often employing measures of relatedness, which inform changes in specialization and explain particular outcomes based on existing specializations (28). These methods of complexity, relatedness, and application of dimensionality reduction are well-served to examine the production and exportation of knowledge. The economic complexity approach is closely aligned with contemporary perspectives in economic geography, which focus on issues such as path dependence, lock-in, and proximity (32, 33). Developments toward evolutionary economic geography (34) recognize the relationship between macro- and micro-level perspectives: in the words of Boschma and Frenken (33), how "spatial structure of the economy emerge from and are transformed by the micro-behaviour of individual and collective agents, and why and how these processes of change are themselves path- and place-dependent".

These approaches can be utilized to examine the degree to which scientific development follows universal patterns conditioned on existing research specializations. For example, Boschma and colleagues (12) applied the principle of relatedness to understand the research topic evolution of cities and showed the emergence of new research topics and the disappearance of existing topics in cities are dependent on their degree of relatedness with existing topics in cities. Guevara and colleagues (11) constructed a research space by using career trajectories of scientists and demonstrated that the research space could predict research evolution of individuals, organizations, and countries. Chinazzi et al. (2019) used an embedding method to predict research evolution in urban areas, providing evidence that the average knowledge density in physics is correlated with scientific and economic development in a country.

The study of economic output with respect to the "product space"—i.e., the network of relatedness between exported products—reveals that the networked structure of industrial advantage is critical to understanding the economic development of nations (31). Following the well-trodden path of Adam Smith, economists argue that division of labor—that is, specialization—is related to economic efficiency; therefore, development is associated with increased capacity and complexity (28, 35). This argument has been explored in regards to economic exports (35); however, less attention has been paid to the relationship between the complexity of scientific exports and economic development, as well as potential universal aspects of these relationships.

In this study, we apply maps of science (30–33) and Revealed Comparative Advantage (RCA) (23)—a common measure for quantifying the economic and production advantages of countries (31)—to examine national science production, by considering scientific disciplines as types of "products" that are exported by countries. That is, we investigate a nation's scientific development through *scientific exports*, in which research articles produced by a country and indexed in international bibliographic databases represent the exported scientific "products" of the nation (21, 40) (see Methods, Supplementary Data). This is an important operationalization for our study; whereas there is a significant amount of scientific production occurs in non-English languages, in grey literature, or governmental reports we argue that it is those works that are made visible through indexation that are the best proxies for *exportation*. This is not to diminish localized scientific activity, but to create a measurement that approximates economic exportation. This allows us to examine how knowledge is constructed within and flows across countries and can be used to inform science policy.

# Results

## *Geography of revealed comparative advantage*

We employ the Revealed Comparative Advantage (RCA) (23) to assess each nation's relative disciplinary strengths based on publications indexed by the Web of Science database (see Methods). If country $c$ produces a greater share of its publications in field $i$ compared to the world average share in the discipline, then $\text{RCA}_{c,i} > 1$ and country $c$ is considered to have a revealed comparative advantage (or specialization) in discipline $i$.

We calculated the RCA for all combinations of 143 disciplines and all countries in our dataset. As expected, the patterns of relative advantage reflect a range of historical, geographical, and cultural factors (see Figure 1). For instance, countries with relative strength in *Botany* are located in tropical areas rich in botanical resources; *Anthropology and Archaeology* features both wealthy and developing nations, reflecting the remnants of colonial science and alluding Basalla's assumption that the science in colonial and post-colonial countries began with Western countries' exploitation of natural resources (23). By contrast, far fewer countries—mostly in North America and Europe—specialize in *Biochemistry & Molecular Biology*, a discipline which requires sufficient funding and sophisticated technologies. Similarly, *Cancer* research is largely concentrated in countries with high cancer mortality (which is associated with longer lifespans) as well as advanced countries with the capacity to invest in clinical research (41). That research and innovation emerge as a response to local issues and threats can also be observed in other contexts. For example, *Agricultural & Food Science* and *Health Policy and Services* are prominent in nations across the global south, where infectious disease (42) and food security (43) are pressing issues. Large emerging economies like China and India are specialized in fields such as *Industrial Engineering* and *Applied Physics* that contribute to industrially-relevant research (3). Similarly, the relative strength of Russia, Ukraine, and Kazakhstan in Applied Physics may be explained as a remnant of the Soviet Union's research priorities (8).

The distribution of disciplinary specialization suggests scientific exportation is affected by geographic, historical, social, and economic factors. Do these idiosyncratic factors dominate the course of scientific development of a nation? Or is there an underlying structure that governs the scientific development of nations?

### *Discipline relatedness network*

Inspired by the relatedness network of economic product exports that underpins national economic development (31) as well as by the studies on scientific research space (10–12), we construct a discipline relatedness network, in which the proximity between disciplines is defined by the minimum conditional probability that two disciplines are co-specialized in a country (see Methods and Supplementary Figure 4-5). The network builds on the idea that disciplines that are co-specialized are likely to require similar knowledge, skills, methods, or equipment. To show its most salient structure, we apply the multi-scale backbone extraction method (44). This "backbone" reveals three clusters—which we confirm in the full network with the Leiden Algorithm (45) (see

Methods)—which we call *Natural*, *Physical*, and *Societal* clusters (see Figure 2a). These clusters——while resembling previous observations (26, 46)—do not conform to the common high-level classifications of disciplines. None of the clusters exclusively coincide with major classifications such as natural sciences, engineering, or medical sciences. The high-level disciplinary classifications appearing in *Natural* cluster (left) are primarily Natural and Medical Sciences. Most disciplines are dependent upon natural resources (e.g., Geology, Entomology, and Agriculture & Food Science), or concern the prevalent medical concerns in low-income areas (e.g., Nutrition & Dietetic and Parasitology). The *Physical* cluster (right) contains primarily physical sciences and engineering, which are commonly considered as foundations for industry-based economic growth (e.g., Chemistry and Applied Physics) and those that require technological investment (e.g., Civil Engineering, Astronomy & Astrophysics, and Aerospace Technology); this cluster suggests the intimate relationships between basic physical science and engineering. The *Societal* cluster (top) is formed by human-centric disciplines that are focused on improving societal welfare, including Medical Sciences (e.g., Psychiatry, Nursing, and Cancer) as well as Social Sciences and Arts & Humanities (e.g., Education, Sociology, and International Relations).

These clusters offer a concise representation of each country's research portfolio. Namely, each country's scientific portfolio can be represented as a point in the simplex of the three clusters (see Figure 2b; see Methods). Aggregating countries based on their income-level classification (47) reveals that niches are largely related to national wealth (see Figure 2b-d). Low-income countries (e.g., Afghanistan, Ethiopia, and South Sudan) tend to be confined to the *Natural* cluster; some of the low-middle countries extend towards the *Physical* disciplines whereas upper-middle income countries are located closer to the center. High-income countries (e.g., United States, France, and Japan) tend to occupy the center and the space between *Natural* and *Societal*, suggesting balanced exportation. This pattern suggests there might be a universal tendency that as a nation's economic power increases, their scientific exports move towards a more balanced portfolio.

### *The principle of relatedness*

To understand the temporal evolution of national research portfolios, we first examine whether the development (or the loss) of revealed comparative advantage follows the principle of relatedness (9–12), which predicts that countries are more likely to develop a new advantage in a discipline that is close to their existing advantages (see Figure 2). By examining the entry (exit) of advantages across each subsequent time step (see Methods), we show the principle of relatedness indeed holds (see Figure 3a-b). The probability of a discipline's entry increases with the density of proximate specialized disciplines ($\beta = 0.33$, 95% CI=[0.30, 0.36]); the probability of a discipline's exit follows the opposite pattern ($\beta = -0.93$, 95% CI=[-0.94,-0.92]).

Moreover, if we aggregate countries based on income groups, we further discover that low-income countries are more strongly constrained by the principle of relatedness than others (one-tailed *t*-test $t_{19}=34$, $P<0.001$) (See Figure 3c-d). In other words, it is more difficult for low-income countries to develop a new relative advantage if it is not in the vicinity of already existing advantage, while wealthy countries are more likely to develop new advantage, thanks to the already existing diverse and complex research portfolio with broader and higher disciplinary production, allowing them to make less portfolio-dependent choices when building scientific capacity. Our result aligns with the recently reported pattern that low-complexity economies experience stronger pull from relatedness and are less likely to enter unrelated activities (48).

The principle of relatedness shapes scientific development, but to what extent? We compare the actual trajectories with a null model that is solely based on the principle of relatedness (see Methods). As shown in Figure 3e, the predicted research profiles converge towards the center of the simplex; in other words, even with the constraining effect of the principle of relatedness, the connections across clusters are strong enough to attract countries towards a balanced research portfolio. By contrast, the aggregated actual trajectories display much weaker attraction towards the center, suggesting scientific development is not entirely dictated by the principle of relatedness but may also be conditioned by the three clusters (see Figure 3e and Supplementary Figure 6-8). The difference is particularly stark for countries specialized in *Natural* cluster, suggesting that low-income countries may face a heavy hurdle breaking into other disciplines (see Supplementary Figure 6-8).

## *Structure of global science*

Meanwhile, global science has been moving from a nested structure to a modular structure. It has been observed that the global economy exhibits a hierarchical (31, 49, 50) (or *nested)* structure, where rich countries can export a wide range of products—especially those that are exported by only a few countries—whereas poor countries can only export a small number of products that can be exported by many (31, 35, 51). This pattern contrasts a more classical theory of specialization, where countries specialize and form a 'modular' structure. Inspired by the tension between these two ideas, we measure the nestedness and modularity of the scientific exports over time (see Methods and Figure 4). In contrast to the case of economic products, we do not observe strong evidence of nestedness; instead, we find that the modularity of the network has been increasing, which is likely associated with the trapping of low-income countries in *Natural* cluster (see Figure 3) and the heavy investment and emphasis of applied sciences in rising economies such as China.

## *Relationships between scientific activities and economic growth*

Motivated by the connection between the economic wealth and their scientific niches and the increasing diversity shown above, we investigate the relationships among scientific diversity, publication volume and economic performance. We measure the diversity of a scientific portfolio with the Gini index of disciplinary RCA values (see Methods). For convenience, we define the *scientific diversity* of a nation as one minus their Gini index. High scientific diversity corresponds to a more balanced and diversified portfolio, whereas low scientific diversity indicates more skewed and specialized exportation.

We find that the number of publications, scientific diversity, GDP and ECI (Economic Complexity Indicator) are all strongly correlated with each other (publication and diversity: Pearson's r=0.91, publication and GDP: Pearson's r=0.92, and publication and ECI: Pearson's r=0.75) (see Figure 5). Over the past 40 years, the average number of publications as well as the average scientific diversity of all income groups have been steadily increasing (see Figure 5d-e). However, this steady growth is not enough to close the gap between income groups, the gap between high-income countries and low-income countries remains wide. Although scientific diversity is correlated with the number of publications, the diversification of research portfolio cannot be explained by the increase in the number of publications alone ($t_{1561}=17.02$, $P<.001$) (see Supplementary Figure 9 for the detail).

Our results with two-way fixed effects panel regression models corroborate a mutual-influence relationship between publication volume and economic development (14, 52, 53) ($t_{704}=2.7$, $P=0.01$, effect size=0.04, 95% CI=[0.01,0.07]) (see Table 1) and ($t_{705}=6.8$, $P<0.001$, effect size=0.29, 95% CI=[0.21,0.38]) (see Supplementary Table 4). As indicated in the model 2 in Table 1, a 10% increase in number of publications is associated with 0.4% relative increase in economic growth rate (see Table 1). The results are robust across different models with GDP and GDP per capita as independent variables respectively ($t_{687}=2.29$, $P=0.03$, effect size=0.02, 95% CI=[0.003,0.04]) (see Table 2). It further shows that, if the publications are divided into the three clusters we identified, the number of publications in *Physical* cluster predicts GDP growth ($t_{695}=1.98$, $P=0.05$, effect size=0.04, 95% CI=[0.001,0.09]) (see Table 1). However, this result is driven by the countries like China that simultaneously exhibited scientific investment focused on Physical cluster and strong economic growth. Removing China from the model makes the coefficient of number of publications in *Physical* cluster statistically insignificant ($t_{688}=0.77$, $P=0.45$, effect size=0.02, 95% CI=[-0.03, 0.06]) (see Supplementary Table 5-6). However, the difference between the coefficient of the number of publication in Physical cluster with all observations and the coefficient of the corresponding parameter excluding China is not significant ($\beta_{\text{with China}}=0.04\pm0.02$, $\beta_{\text{without}}$

$_{China}$=0.02±0.02) (54). Although we cannot exclude the possibility that China's focused investment in Physical cluster might have played an important role in its economic growth, our models do not provide enough evidence to demonstrate whether the strength in Physical cluster is associated with the future economic growth across countries. The relationship between scientific development and economic development may be contingent on countries and complex.

In contrast to the theory that diversity and complexity of economy are closely linked to the economic growth, we could not find strong associations between the diversity of scientific portfolio and economic growth. Scientific diversity could neither predict GDP growth nor predict the growth of publication ($t_{704}$=-0.78, P=0.44, effect size=-0.08, 95% CI=[-0.27,0.11]) (see Table 1) and ($t_{704}$ =-0.77, P=0.44, effect size=-0.10, 95% CI=[-0.36,0.16]) (see Supplementary Table 4), whereas GDP predicts the growth rate of scientific diversity ($t_{705}$=2.71, P=0.007, effect size=0.43, 95% CI=[0.12,0.74]) (see Supplementary Table 7). Our results suggest that a balanced research portfolio may be a result, rather than the cause of economic development. However, scientific diversity is negatively associated with the similarity between newly entered disciplines and the existing advantages, suggesting that balanced research profile may be associated with more flexibility to develop research areas ($t_{703}$=-1.71, P=0.09, effect size=-0.60, 95% CI=[-1.29,0.09]) (see Table 3). Countries with high diversity tend to develop more easily beyond their current research advantages.

## Discussion

It is widely believed that scientific development holds the key to the future prosperity of a nation

(55, 56). Yet, whether there are universal structural patterns of scientific development at the national level remains an open question. By analyzing more than 30 million scientific publications across 217 countries spanning the period 1973-2017, we provide a large-scale temporal analysis of national science development. We find that the disciplinary proximity network constructed from these publications exhibits three clusters of disciplines which roughly capture the relative advantages of countries across the spectrum of economic wealth. Although each country's position in the network is shaped by various historical, geographical, social, and economic factors, the three-cluster structure still conditions their scientific development. We further reveal that, although individual country is moving towards a more balanced research profile, global science is becoming more modular. We also confirm that the economic wealth and scientific publication are mutually predictive of each other, suggesting a strong feedback loop. Finally, we find evidence that the economic growth leads to higher scientific diversity and countries with diverse research portfolios are more flexible to develop new research areas.

Our results in part reaffirm the general patterns observed in previous studies on the structure of knowledge space and the principle of relatedness (10–12, 22, 23, 26, 46, 57). The clusters and the niches that are occupied by nations show some semblance of Comte's "Hierarchy of the Sciences" (1855) hypothesis—that science progresses from natural sciences that require readily-available simple subjects, towards social sciences that deal with more complex subjects. At the same time, the prominence of *Natural* disciplines in low-income countries resonates with Basalla's "Spread of Western Science" (Basalla, n.d.), pointing to the colonial exploitation of natural resources. Klavans and Boyack (2017) and Moya-Anegón and Herrero-Solana (2013) identified similar disciplinary clusters that condition on national scientific development. Furthermore, studies have demonstrated the principle of relatedness in knowledge domain (9–12). Building on this literature, by taking a country-level approach and network analysis, our work quantitatively identifies three disciplinary clusters where country's niches in the knowledge space is associated with its economic development. We further reveal that the principle of relatedness alone is not enough to explain the dynamic evolution of research areas in countries.

This study has several limitations. First, it relies on a bibliographic database created and maintained by a Western scientific enterprise. Therefore, it overestimates research from western countries and publications in English while underestimating the production in other nations and languages (see Supplementary Information). Still, we argue that our operationalization is reasonable under the analogy to product exportation (31, 35) and the status of English as the de-facto *lingua franca* (58) of science. Second, many of our analyses considered the RCA matrix as a bipartite network. This approximation is not strictly valid because the edges are not independent from each other. Finally, our regression models are not free from multicollinearity issues as there exist significant correlation between GDP and the number of publications. Moreover, we show that the inferred relationships between scientific enterprise and economic growth can be easily driven by a small number of countries by demonstrating that the association between the publications in Physical cluster and economic growth is driven by China, which have achieved strong growth in both scientific production in Physical cluster and economy. Due to the complexity of economic and scientific development, we note that reliable causal inference with country-level data is often infeasible, and our results do not necessarily confirm nor reject a direct causal relationship between national scientific development and economic growth. There exist many unobserved hidden confounders and complex feedback mechanisms between scientific and economic development.

Despite those limitations, our empirical framework may provide a useful perspective to study the structure and evolution of national scientific portfolios and the relationship to economic development. Our results call for attention to the barriers faced by low-income countries in building their scientific capacity, and the potential consequences on future scientific capacity and economic growth. Our results also highlight the importance of considering scientific capacity in the study of economic development. We hope our analysis opens a new avenue towards the understanding of the mechanisms of scientific development as well as its relationship to economic prosperity.

## Methods

**Data.** The dataset was drawn from the Clarivate Analytics' Web of Science database hosted and managed by the *Observatoire des Sciences et des Technologies* at the University of Montreal. The Web of Science database contains three main citation indices: The Science Citation Index Expanded, the Social Science Citation Index, and the Arts and Humanities Citation Index. We used all indexed publication records listed as being published between 1973 to 2017, which included 37,479,532 papers published across 20,252 scholarly journals. To examine temporal patterns, we split the data into nine five-year snapshots. We limited this set to only journal articles, review articles, and notes (discontinued in 1991 but included in articles). We also excluded any publication that did not list any institutional address, and publications that could not be assigned a disciplinary category according to the steps below. After these filters, the dataset contained 35,793,320 papers published across 20,123 scholarly journals (See Figure S2).

Discipline classification of publications is based on the National Science Foundation typology of journals, which categorizes papers into a hierarchy of disciplines. The high-level and granular classification was further complemented with an in-house classification of the Arts and Humanities (59). The resulting classification scheme contains 144 granular categories. After removing "Unknown" from the 144 granular categories, we manually classified each of the 143 categories into one of five broad categories: "Natural Science", "Medical Science", "Engineering", "Social Science", and "Arts and Humanities"; this scheme is used to color nodes in Figure 2.

Publications are associated with nations using the institutional addresses listed by the authors. We assign a full unit credit of a publication to every country of affiliation represented on the paper's author byline ("full counting"). For example, a paper listing five authors—two with affiliations in the United States, two in Canada, and one in the Netherlands—would count as one paper to all three countries. Full counting method assumes each author contributes equally to the publication. Fractional counting and counting based on corresponding authorship are another two widely used counting methods. These counting methods are highly correlated at the macro level (60). However, Web of Science has a highly inaccurate coverage on the corresponding author information before 2008, where corresponding author is, in most cases, simply assigned to the first author/institution. Given the diachronic nature of our analysis, we were unable to utilize counting method based on corresponding authorship. The discipline network constructed from fractional counting share high similarity with network constructed using full counting method. See Supplementary Information's data section for more details.

We use data on national GDP from the World Bank (47, 61) to approximate the economic wealth of each country. The dataset covers 264 countries from 1960 to 2019. Income classification comes from World bank database (47) which contains 224 countries between 1987 and 2018. We convert the annual classification to a time snapshot classification by assigning each country to its most frequent income group during each period. See Supplementary Information's data section for more details.

**Revealed Comparative Advantage.** The revealed comparative advantage (RCA) of country $c$ in discipline $i$ is defined as:

$$\text{RCA}_{c,i} = \frac{\mathcal{P}(c,i)/\sum_i \mathcal{P}(c,i)}{\sum_c \mathcal{P}(c,i)/\sum_{c,i} \mathcal{P}(c,i)}$$

where $\mathcal{P}(c,i)$ is the number of publications produced and "exported"—the number of publications which is indexed in the Web of Science—by country $c$ in discipline $i$, $\sum_i \mathcal{P}(c,i)$ is the total number of publications produced by country $c$, $\sum_c \mathcal{P}(c,i)$ is the total number of publications produced in a discipline globally, and $\sum_{c,i} \mathcal{P}(c,i)$ is the total number of publications across all countries and disciplines.

**Disciplinary Proximity.** The proximity between disciplines $i$ and $j$ is defined as the minimum of the pairwise conditional probabilities of a country having an advantage (RCA > 1) in one discipline given an advantage in another:

$$\phi_{ij} = \min\{P(\text{RCA}_i > 1 | \text{RCA}_j > 1), P(\text{RCA}_j > 1 | \text{RCA}_i > 1)\}$$

$\phi$ is a $143 \times 143$ matrix that captures the proximity between pairs of disciplines (see Supplementary Figure 4).

**Identifying the disciplinary clusters.** The relatedness network is constructed from the disciplinary proximity matrix derived from aggregating data across all years (from 1973 to 2017). The network is fixed over the analysis. Although the network structure changes over time, networks derived from a snapshot of data closely resemble with the aggregated network (see Supplementary Figure 5). The multi-scale backbone extraction method (44) exposes three visual clusters when laid out with a force-directed layout algorithm (Gephi's ForceAtlas2). We formally confirm the structure by applying a community detection algorithm (the Leiden algorithm (45)) to the full network. We use modularity as the quality function and ran the model 50 times to obtain consensus. Other methods produce similar results although some partitions the network into smaller communities (see Supplementary Information's Disciplinary Relatedness Network section).

**Position within the simplex.** Position within the simplex measure countries' cluster-level specialization concentration. We first calculate $C_i = n_i/N_i$, where $n_i$ is the number of disciplines in cluster $i$ with RCA > 1, $N_i$ is the total number of disciplines in cluster $i$. Then we normalize $C_i$ so that $\sum_i C_i = 1$.

**The Density of existing advantages and the null development model.** The density of existing advantages around a given discipline is defined as follows:

$$\omega_j^k = \frac{\sum_i x_i \phi_{ij}}{\sum_i \phi_{ij}}$$

where $\phi_{ij}$ is the proximity between discipline $i$ and $j$, and $x_i = 1$ if $RCA_{ki} > 1$ else $x_i = 0$, and the density of existing advantages, $\omega_j^k$, is the proximity-weighted sum of all disciplines that are connected to $j$ with $RCA_{ki} > 1$. We bin the density values and aggregate across countries and time periods to calculate the probability of entry and exit, given the density. We also perform a bootstrap sampling with 20 samples to estimate the uncertainty of the slope and report the mean and standard deviation of the slopes across bootstrapped samples. A linear regression model (OLS) is fit by pooling all bootstrap samples to obtain the parameters (intercept and slope) for the null model. The null model works as following: for every inactive (RCA < 1) discipline, we assign a probability the discipline will be entered (RCA > 1) in the subsequent time period based on its current density using the intercept and slope obtained from the pooled regression model that include all countries. We use the same procedure for the exit. For each time period and each country, the new entered and exited disciplines are sampled using the null model while preserving the number of new entered and exited disciplines in the next time period. We repeat this procedure 100 times. When visualizing the actual profile and the predicted profile on the simplex, to reduce the influence of extreme cases, we remove data points located on the boundary of the simplex. To smooth out the noise, we aggregate data points within each rhombus with the side length of 0.1 that tessellates across the simplex. We observe that the difference between actual trajectory and the predicted trajectory is robust against the direction of rhombus.

**Modularity and Nestedness.** We use the country-discipline bipartite network to represent knowledge exportation. Country $c$ is connected to discipline $i$ if $RCA_{c,i} > 1$. Modularity (62) of the country-discipline bipartite network is defined as:

$$Q = \frac{1}{m}\sum_{i=1}^{p}\sum_{j=1}^{q}(A_{ij} - P_{ij})\delta(g_i, h_j)$$

Where $m$ is the number of links, $A_{ij}$ equals to 1 if there is a link from node $i$ to node $j$, $P_{ij}$ is the probability the edge between $i$ and $j$ exists under the null model, $g_i$ and $h_j$ are communities that the country and discipline belong to. The community of a country is decided by its largest cluster level revealed comparative advantage; for example, China is classified to *Physical* cluster since it has highest cluster-level RCA value in *Physical* cluster. The community of disciplines is defined by the Leiden algorithm. Although the elements of the RCA matrix are not strictly independent from each other, we use $P_{ij} = \frac{k_i d_j}{m}$ (where $k_i$ and $d_j$ are the degree of node $i$ and $j$ respectively) as an approximation. Larger modularity means countries tend to be specialized in one of the three clusters rather than having advantages spread across multiple clusters.

Nestedness is measured by the overlap and decreasing fill (NODF) (63) method. NODF

measures the degree of overlapping between row pairs and column pairs in the adjacency matrix. The metric is defined as

$$NODF = \frac{\sum N_{paired}}{\left[\frac{n(n-1)}{2}\right] + \left[\frac{m(m-1)}{2}\right]}$$

Where $\sum N_{paired}$ is the averaged degree of nestedness for each pair of row and column based on the principles of decreasing fill and paired overlap (63), *n* and *m* are the number of rows and columns.

We use a null model to test whether modularity and nestedness are significant. We construct the null model of the bipartite network by swapping edges between node pairs while constraining the degree of each node which we refer to as the Fixed-Fixed null model.

**Scientific Diversity.** The Gini index of a nation's RCA values across disciplines is used to capture the scientific diversity of a nation. For convenience, we use 1 minus the Gini index as a measure of scientific diversity. If all disciplines have the same RCA value in the country, the diversity value would be 1. If a country only produces scientific publications in one discipline, then the diversity value would be 0. To investigate the dynamic relationship between scientific diversity and economic power, we project countries evolution into the diversity-GDP plane. To smooth out noise, we averaged the trajectory in each grid with width equals to 0.1 and height equals to 0.5. The starting point of arrow represents average of all displacements whose starting points were in the grid. The direction and length of arrows are computed by averaging the subsequent displacements of all countries within a grid.

**Regression Analysis**: We use fixed-effect panel regression model to investigate the relationship between economic growth and scientific development. The model is written as following:

$$Y_{c,t} = \beta_0 + \beta_1 X_{1,ct} + \beta_2 X_{2,ct} + \cdots + \beta_k X_{k,ct} + \alpha_c + \alpha_t$$

Where $c$ denotes countries, $t$ denotes time periods, $Y_{c,t}$ is the investigated dependent variable, $\alpha_c$ and $\alpha_t$ are the country-specific and time specific intercepts that capture the heterogeneity across countries and across time periods. The dependent variables involved in our analysis are log ratio of GDP growth rate, log ratio of publication growth rate, scientific diversity growth rate and averaged disciplinary similarity. Growth rate is measured as $N_{c,t+1}/N_{c,t}$ where t+1 represents the next time period which is the following 5 years. The included controlled variables are averaged GDP value, averaged ECI value and averaged number of populations. The investigated independent variables are the total number of publications, scientific diversity, and the number of publications in *Natural*, *Physical* and *Societal* clusters. We apply log-transformation with base 10 to GDP, GDP per capita, the number of publications the number of populations and the growth rate.

Averaged disciplinary similarity measures how similar the new entered disciplines are comparing with the current existing advantages. The averaged disciplinary similarity is defined as:

$$\rho_c = \sum_{i}^{n} \frac{\omega_i^c}{n}$$

Where $n$ is the number of new entered disciplines and $\omega_i^c$ is the normalized density of new entered disciplines. Normalized density is measured as the z-score of raw disciplinary density of all disadvantaged disciplines. High averaged disciplinary similarity indicates the new entered disciplines have higher similarity with existing advantaged disciplines compared with average similarity between advantaged disciplines and inactivated disciplines.

**Data Availability**: Data used in this study is available at

https://figshare.com/articles/journal_contribution/Untitled_Item/13623035/3

**Code Availability**: The code used for data processing and analysis will be available at

https://github.com/yy/national-science-exports

**Acknowledgements**

L.M., D.M., and Y.Y.A acknowledges funding support from the Air Force Office of Scientific Research under award number FA9550-19-1-0391. The funder had no role in study design, data collection and analysis, decision to publish or preparation of the manuscript. We thank Staša Milojević, Jisung Yoon, Sadamori Kojaku, Byungkyu Lee, Bruce Weinberg, Paula Stephan, Santiago Schnell, Tao Zhou, Junming Huang, Jian Gao, Yi Bu, Qianming Zhang, Brian Uzzi, Hyejin Youn, and Nima Dehmamy for helpful discussion and comments.

**Author Contributions:**

L.M. and D.M. conceived the study; All authors contributed to the design of the study; V.L. prepared the primary datasets; L.M., D.M., V.L., Y.Y.A performed analysis; All authors contributed to the interpretation of the results and writing of the manuscript.

**Competing Interests**

The authors declare no competing interests.

**Tables**

Table 1 Regression results of predicting growth rate of GDP

| | *Dependent variable:* |
|---|---|
| | GDP growth (log-ratio) |

|  | (1) | (2) | (3) | (4) | (5) | (6) | (7) |
|---|---|---|---|---|---|---|---|
| Log GDP | -0.42*** | -0.44*** | -0.43*** | -0.43*** | -0.42*** | -0.43*** | -0.44*** |
|  | (-0.48, -0.36) | (-0.50, -0.38) | (-0.49, -0.37) | (-0.49, -0.37) | (-0.48, -0.36) | (-0.49, -0.37) | (-0.50, -0.38) |
|  | p = 0.00 | p = 0.00 | p = 0.00 | p = 0.00 | p = 0.00 | p = 0.00 | p = 0.00 |
| ECI | 0.01 | 0.003 | 0.004 | 0.01 | 0.01 | 0.01 | 0.002 |
|  | (-0.02, 0.04) | (-0.03, 0.03) | (-0.03, 0.03) | (-0.03, 0.04) | (-0.02, 0.04) | (-0.03, 0.04) | (-0.03, 0.03) |
|  | p = 0.64 | p = 0.86 | p = 0.79 | p = 0.73 | p = 0.65 | p = 0.72 | p = 0.92 |
| Log Population | 0.15** | 0.11* | 0.12** | 0.13** | 0.15** | 0.13** |  |
|  | (0.04, 0.27) | (-0.01, 0.23) | (0.002, 0.25) | (0.003, 0.25) | (0.02, 0.27) | (0.004, 0.25) |  |
|  | p = 0.02 | p = 0.09 | p = 0.05 | p = 0.05 | p = 0.03 | p = 0.05 |  |
| Log no.Pub |  | 0.04*** |  |  |  |  | 0.06*** |
|  |  | (0.01, 0.07) |  |  |  |  | (0.02, 0.09) |
|  |  | p = 0.01 |  |  |  |  | p = 0.002 |
| Log no.Natural |  |  | 0.03** |  |  | -0.01 |  |
|  |  |  | (0.004, 0.07) |  |  | (-0.05, 0.04) |  |
|  |  |  | p = 0.03 |  |  | p = 0.84 |  |
| Log no.Physical |  |  |  | 0.04*** |  | 0.04** |  |
|  |  |  |  | (0.01, 0.07) |  | (0.001, 0.09) |  |
|  |  |  |  | p = 0.005 |  | p = 0.05 |  |
| Log no.Societal |  |  |  |  | 0.03* |  |  |
|  |  |  |  |  | (-0.003, 0.06) |  |  |
|  |  |  |  |  | p = 0.09 |  |  |
| Diversity |  |  |  |  |  |  | -0.08 |
|  |  |  |  |  |  |  | (-0.27, 0.11) |
|  |  |  |  |  |  |  | p = 0.44 |
| Observations | 836 | 836 | 836 | 828 | 827 | 828 | 836 |
| $R^2$ | 0.21 | 0.22 | 0.22 | 0.22 | 0.21 | 0.22 | 0.22 |
| Adjusted $R^2$ | 0.07 | 0.08 | 0.07 | 0.07 | 0.06 | 0.07 | 0.07 |
| F Statistic | 63.57*** (df = 3; 705) | 50.00*** (df = 4; 704) | 49.19*** (df = 4; 704) | 48.32*** (df = 4; 696) | 46.04*** (df = 4; 695) | 38.61*** (df = 5; 695) | 49.23*** (df = 4; 704) |

*Note:* The P-value is derived from a two-sided t-test. *p<0.1;**p<0.05;***p<0.01.

Table 2 Regression results of predicting growth rate of GDP per capita

|  | *Dependent variable:* | | | | | | |
|---|---|---|---|---|---|---|---|
|  | GDP per capita growth (log-ratio) | | | | | | |
|  | (1) | (2) | (3) | (4) | (5) | (6) | (7) |
| Log GDP per capita | -0.25*** | -0.27*** | -0.26*** | -0.27*** | -0.26*** | -0.27*** | -0.22*** |

|  | (1) | (2) | (3) | (4) | (5) | (6) | (7) |
|---|---|---|---|---|---|---|---|
|  | (-0.29, -0.21) | (-0.32, -0.22) | (-0.31, -0.22) | (-0.32, -0.23) | (-0.31, -0.21) | (-0.32, -0.23) | (-0.26, -0.18) |
|  | p = 0.00 | p = 0.00 | p = 0.00 | p = 0.00 | p = 0.00 | p = 0.00 | p = 0.00 |
| ECI | 0.01* | 0.01 | 0.01 | 0.01 | 0.01 | 0.01 | 0.01* |
|  | (-0.002, 0.03) | (-0.01, 0.03) | (-0.004, 0.03) | (-0.005, 0.03) | (-0.003, 0.03) | (-0.005, 0.03) | (-0.002, 0.03) |
|  | p = 0.10 | p = 0.19 | p = 0.15 | p = 0.16 | p = 0.11 | p = 0.16 | p = 0.10 |
| Log Population | -0.11*** | -0.14*** | -0.13*** | -0.15*** | -0.12*** | -0.15*** |  |
|  | (-0.18, -0.04) | (-0.22, -0.07) | (-0.21, -0.06) | (-0.22, -0.07) | (-0.20, -0.04) | (-0.22, -0.07) |  |
|  | p = 0.003 | p = 0.0003 | p = 0.001 | p = 0.0002 | p = 0.003 | p = 0.0002 |  |
| Log no.Pub |  | 0.02** |  |  |  |  | 0.02** |
|  |  | (0.003, 0.04) |  |  |  |  | (0.001, 0.04) |
|  |  | p = 0.03 |  |  |  |  | p = 0.04 |
| Log no.Natural |  |  | 0.02* |  |  | -0.003 |  |
|  |  |  | (-0.001, 0.03) |  |  | (-0.03, 0.02) |  |
|  |  |  | p = 0.08 |  |  | p = 0.81 |  |
| Log no.Physical |  |  |  | 0.02** |  | 0.02* |  |
|  |  |  |  | (0.003, 0.03) |  | (-0.002, 0.04) |  |
|  |  |  |  | p = 0.03 |  | p = 0.09 |  |
| Log no.Societal |  |  |  |  | 0.01 |  |  |
|  |  |  |  |  | (-0.01, 0.02) |  |  |
|  |  |  |  |  | p = 0.44 |  |  |
| Diversity |  |  |  |  |  |  | -0.12** |
|  |  |  |  |  |  |  | (-0.22, -0.02) |
|  |  |  |  |  |  |  | p = 0.03 |
| Observations | 818 | 818 | 818 | 811 | 810 | 811 | 818 |
| $R^2$ | 0.16 | 0.17 | 0.17 | 0.17 | 0.16 | 0.17 | 0.16 |
| Adjusted $R^2$ | 0.01 | 0.01 | 0.01 | 0.01 | 0.005 | 0.01 | 0.001 |
| F Statistic | 44.60*** (df = 3; 688) | 34.97*** (df = 4; 687) | 34.35*** (df = 4; 687) | 35.45*** (df = 4; 680) | 33.43*** (df = 4; 679) | 28.33*** (df = 5; 679) | 32.61*** (df = 4; 687) |

*Note:* The P-value is derived from a two-sided t-test. *p<0.1; **p<0.05; ***p<0.01

Table 3 Regression results of predicting average similarity of new entered disciplines

|  | Dependent variable: | |
|---|---|---|
|  | Similarity | |
|  | (1) | (2) |
| Log GDP | -0.11 | -0.09 |
|  | (-0.34, 0.11) | (-0.31, 0.14) |
|  | p = 0.34 | p = 0.45 |
| ECI | -0.02 | -0.02 |
|  | (-0.13, 0.09) | (-0.13, 0.09) |
|  | p = 0.70 | p = 0.76 |
| Log no.Pub | -0.02 | 0.04 |
|  | (-0.13, 0.09) | (-0.09, 0.17) |
|  | p = 0.70 | p = 0.57 |
| Diversity |  | -0.60* |
|  |  | (-1.29, 0.09) |
|  |  | p = 0.09 |
| Observations | 835 | 835 |
| $R^2$ | 0.002 | 0.01 |
| Adjusted $R^2$ | -0.18 | -0.18 |
| F Statistic | 0.54 (df = 3; 704) | 1.14 (df = 4; 703) |

Note: The P-value is derived from a two-sided t-test. *p<0.1; **p<0.05; ***p<0.01

**Figure**

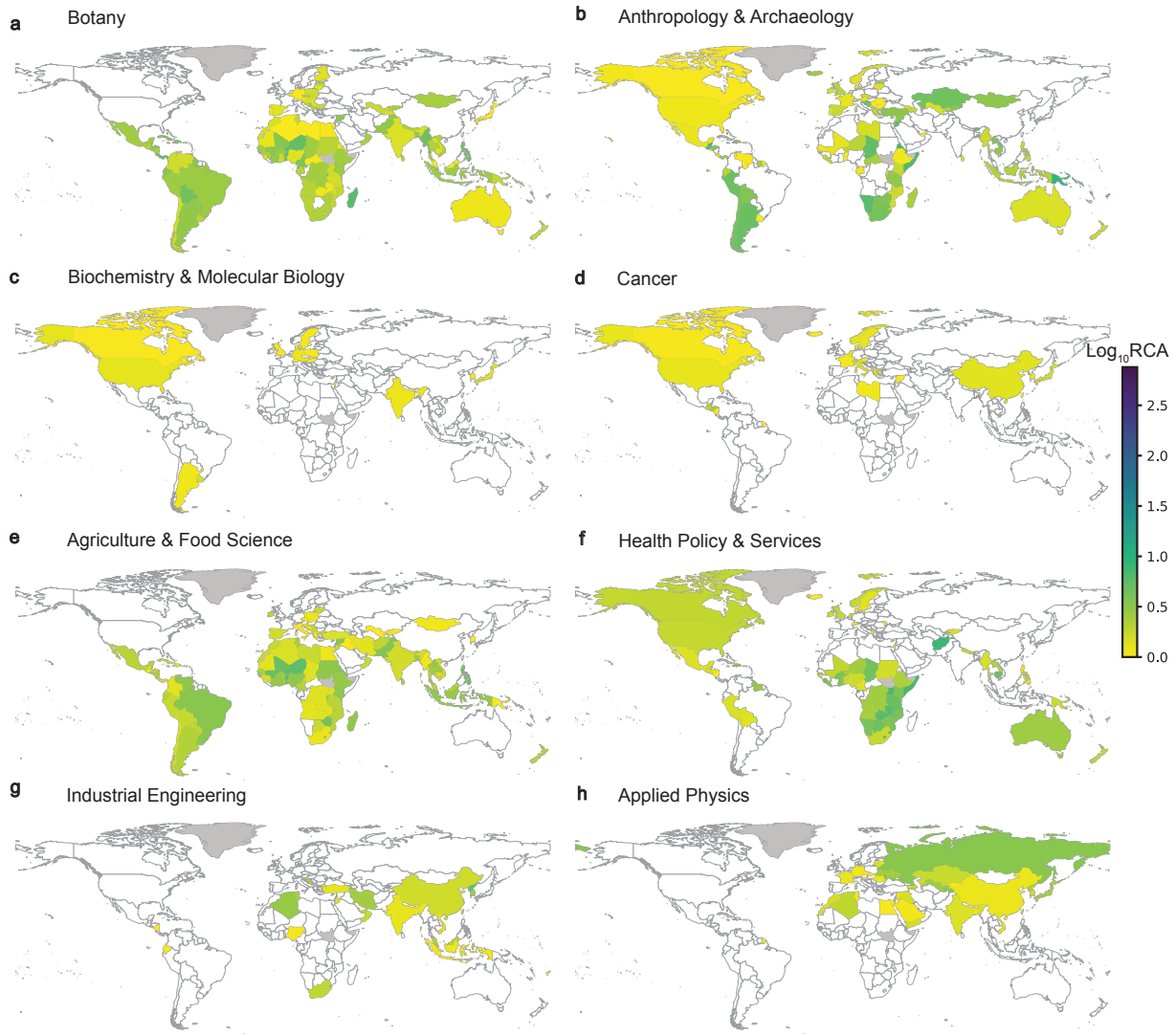

***Figure 1. Disciplinary specialization reflects geographical, historical and economic factors.***
*Eight examples illustrate the distribution of disciplinary specializations. Discipline specialization is measured by Revealed Comparative Advantage (RCA). Color represents the Logarithm of RCA; a nation is only colored if its $\log_{10} \text{RCA}_{c,i} > 0$. Grey corresponds to nations that were not represented in our dataset. Botany, Anthropology, and Archaeology reflect the presence and access to natural and anthropological resources in a country. Economic inequality underpin specialization in resource-intensive disciplines like Biochemistry & Molecular Biology and Cancer. Local issues also drive research, as can be seen from the distribution of Agricultural & Food Science and Health Policy & Services. The distribution of Industrial Engineering and Applied Physics likely reflects national economic priorities and policies. The map data is downloaded from Natural Earth. Free vector and raster map data @ naturalearthdata.com.*

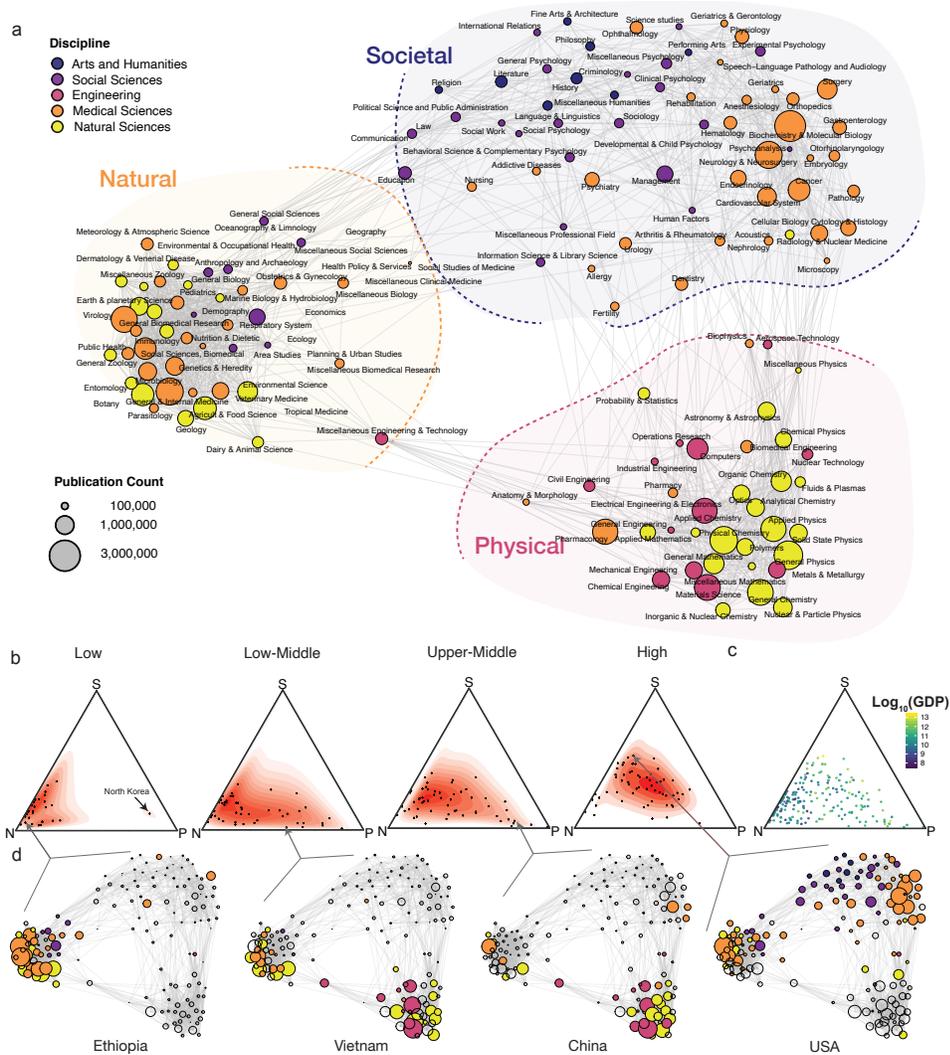

*Figure 2. The Structure of the disciplinary proximity network and national development. (a) The backbone of the disciplinary relatedness network reveals three clusters, which we call Natural, Physical, and Societal. Each node corresponds to a discipline and the weight of an edge captures the minimum conditional probability of co-specialization (see Methods). The area of a node is proportional to the number of total publications indexed in that discipline. Node color maps to five broad disciplinary categories. (b) Nations are classified into four groups by their income level: Low, Low-Middle, Upper-Middle and High (from left to right). Dots correspond to nations, and a nation's position inside the simplex is calculated as the fraction of advantaged disciplines in each cluster normalized by its total number of advantaged disciplines. The density estimate of each income group is shown in red. (c) National research profile snapshots (2013-2017) and GDP. Points are colored according to the nation's log-transformed GDP. (d) Four example countries, Ethiopia, Vietnam, China, and the United States (USA) at 2013-2017. Only the discipline with an advantage ($\log_{10} RCA > 0$) are colored. Node colors are the same as in (a).*

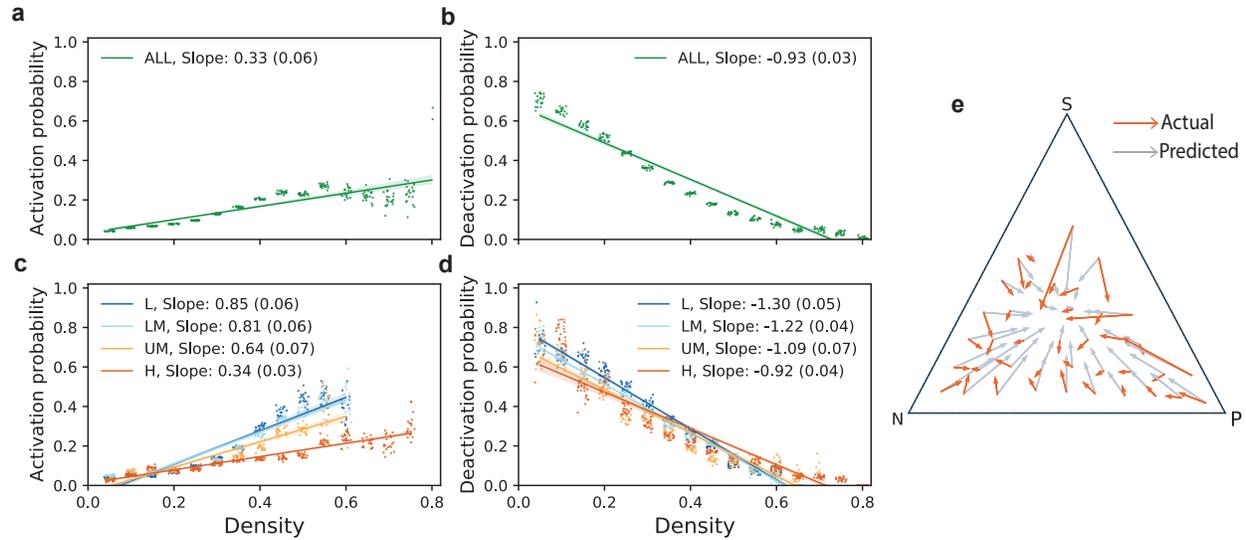

*Figure 3. The principle of relatedness dictates the development and loss of competencies (a) Probability of a new relative advantage in the next time period given the density of existing advantages surrounding the discipline. Dots represent the estimated probabilities from Bootstrapping and solid lines are the estimated regression line from bootstrapped samples. Translucent band lines and the number in the parentheses describe bootstrapped 95% confidence interval and the standard deviation of the estimated slope. 211 countries are involved in the analysis and we performed 20 times of bootstrapping. (b) Probability that an advantaged discipline will lose its advantage in the next period given the density of existing advantages. (c)-(d) Same plots where countries are grouped based on their income class. Low-income group contains 72 countries; low-middle group contains 104 countries; upper-middle group contains 74 countries and high-income contains 56 countries. Because countries transit to different income group during different time period. Therefore countries are double-counted in the abovementioned numbers. (e) We show the predicted and actual evolution in the simplex. Arrows point to the average simplex position of countries in the next period. Red arrows represent the empirical movement while gray arrows represent the movement predicted from the null model based only on the principle of relatedness.*

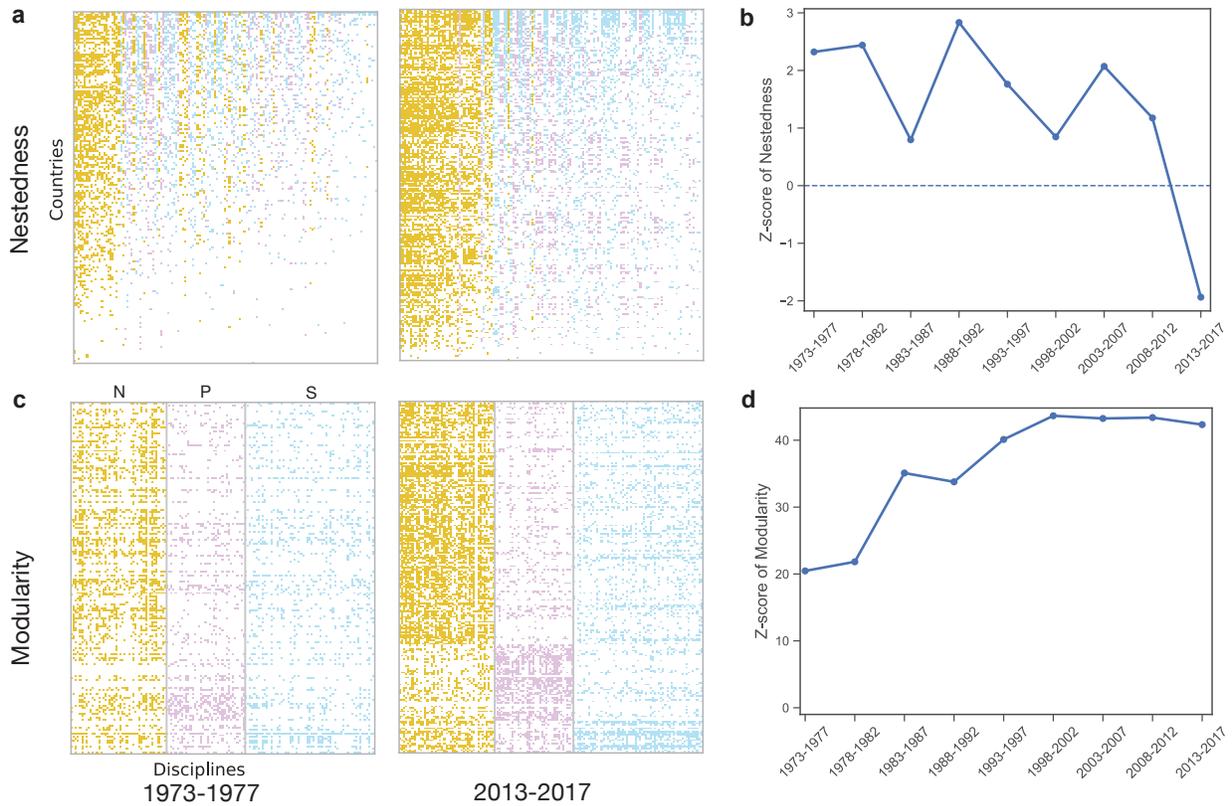

*Figure 4. Nestedness and modularity of global science* (a) The country-discipline RCA matrices of the earliest and the most recent periods where rows and columns are arranged in a descending order of number of advantages. (b) The z-score of nestedness over time which is calculated through Fixed-Fixed null model. (c) The country-discipline RCA matrices of the earliest and the most recent time periods where disciplines(columns) are arranged by its classification into the three clusters. (d) The z-score of modularity over time which shares the same null matrix as used for calculating nestedness.

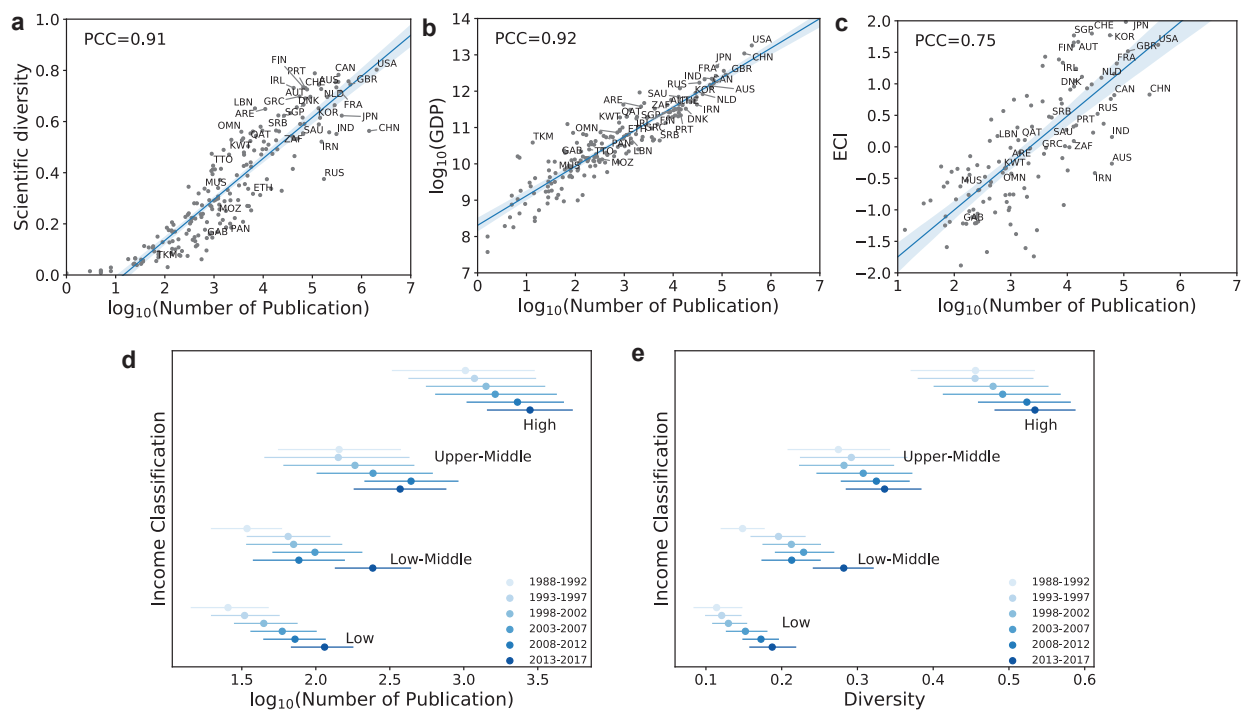

*Figure 5. Scientific production is correlated with national development indicators. (a) Number of publications is strongly correlated with scientific diversity (defined as one minus the GINI index of the RCA values of a country). Lines represent a linear regression model fit with x-axis variable as the independent variable and y-axis variable as dependent variable. Translucent band lines describe bootstrapped 95% confidence interval. PCC stands for Pearson correlation coefficient. (b)-(c) The relationship between scientific publication volume and nation's GDP (c) and their economic complexity index. (d)-(e) The temporal development of (d) the number of productions and (e) scientific diversity by income group. The point shows the mean value of each group. Error bars represent the 95% confidence interval of the mean value drawn from bootstrapping. The number of countries in income group during period is presented in Supplementary Table S3. 1000 times of iterations are used to compute the confidence interval. Point represents the mean value. Error bars represent the 95% confidence interval drawn from bootstrapping. The number of countries in each time period is presented in Supplementary Table S2.*

# References

1. National Science Board, National Science Foundation. 2019, "Publications Output: U.S. Trends and International Comparisons" (2019).

2. J. Tollefson, China declared world's largest producer of scientific articles. *Nature* **553**, 390–390 (2018).

3. P. Zhou, L. Leydesdorff, The emergence of China as a leading nation in science. *Res. Policy* **35**, 83–104 (2006).

4. R. V. Noorden, Science in East Asia — by the numbers. *Nature* **558**, 500–501 (2018).

5. K.-W. Li, *Capitalist Development and Economism in East Asia: The Rise of Hong Kong, Singapore, Taiwan and South Korea.* (Routledge, 2002).

6. D. N. Livingstone, *Putting Science in Its Place: Geographies of Scientific Knowledge* (University of Chicago Press, 2010).

7. S. Seth, Putting knowledge in its place: science, colonialism, and the postcolonial. *Postcolonial Stud.* **12**, 373–388 (2009).

8. J. Kozlowski, S. Radosevic, D. Ircha, History matters: The inherited disciplinary structure of the post-communist science in countries of central and eastern Europe and its restructuring. *Scientometrics* **45**, 137–166 (1999).

9. C. A. Hidalgo, *et al.*, "The Principle of Relatedness" in *Unifying Themes in Complex Systems IX*, Springer Proceedings in Complexity., A. J. Morales, C. Gershenson, D. Braha, A. A. Minai, Y. Bar-Yam, Eds. (Springer International Publishing, 2018), pp. 451–457.

10. M. Chinazzi, B. Gonçalves, Q. Zhang, A. Vespignani, Mapping the physics research space: a machine learning approach. *EPJ Data Sci.* **8**, 33 (2019).

11. M. R. Guevara, D. Hartmann, M. Aristarán, M. Mendoza, C. A. Hidalgo, The research space: using career paths to predict the evolution of the research output of individuals, institutions, and nations. *Scientometrics* **109**, 1695–1709 (2016).

12. R. Boschma, G. Heimeriks, P.-A. Balland, Scientific knowledge dynamics and relatedness in biotech cities. *Res. Policy* **43**, 107–114 (2014).

13. S. Paula, *How Economics Shapes Science* (Harvard University Press, 2012).

14. L.-C. Lee, P.-H. Lin, Y.-W. Chuang, Y.-Y. Lee, Research output and economic productivity: a Granger causality test. *Scientometrics* **89**, 465–478 (2011).

15. R. R. Kumar, P. J. Stauvermann, A. Patel, Exploring the link between research and economic growth: an empirical study of China and USA. *Qual. Quant.* **50**, 1073–1091 (2016).

16. T. Hornyak, Chilean research grows despite poor investment. *Nat. Index* (2016).

17. L. Bronfman, A panorama of Chilean astronomy. *The Messenger* **107**, 14–18 (2002).

18. A. Bajak, Chile's chance to embrace science for the twenty-first century. *Nature* **552**, S53–

S55 (2017).

19. H. W. Yeom, South Korean science needs restructuring. *Nature* **558**, 511–513 (2018).

20. National Science Board, "Science and Engineering Indicators 2018" (2018).

21. R. M. May, The Scientific Wealth of Nations. *Science* **275**, 793–796 (1997).

22. A. Comte, *The Positive Philosophy of Auguste Comte* (C. Blanchard, 1855).

23. G. Basalla, The Spread of Western Science. *Science* **156**, 611–622 (1967).

24. W. Anderson, Remembering the Spread of Western Science. *Hist. Rec. Aust. Sci.* **29**, 73 (2018).

25. D. Raina, From west to non-west? Basalla's three-stage model revisited. *Sci. Cult.* **8**, 497–516 (1999).

26. F. Moya-Anegón, V. Herrero-Solana, Worldwide Topology of the Scientific Subject Profile: A Macro Approach in the Country Level. *PLoS ONE* **8**, e83222 (2013).

27. G. Cimini, A. Gabrielli, F. S. Labini, The Scientific Competitiveness of Nations. *PLOS ONE* **9**, e113470 (2014).

28. C. A. Hidalgo, Economic complexity theory and applications. *Nat. Rev. Phys.* **3**, 92–113 (2021).

29. I. Hong, M. R. Frank, I. Rahwan, W.-S. Jung, H. Youn, The universal pathway to innovative urban economies. *Sci. Adv.* **6**, eaba4934 (2020).

30. A. Alabdulkareem, *et al.*, Unpacking the polarization of workplace skills. *Sci. Adv.* **4**, eaao6030 (2018).

31. C. A. Hidalgo, B. Klinger, A.-L. Barabasi, R. Hausmann, The Product Space Conditions the Development of Nations. *Science* **317**, 482–487 (2007).

32. R. A. Boschma, F. Koen, Evolutionary economics and industry location. *Rev. Reg. Res.* **23**, 183–200 (2003).

33. R. A. Boschma, K. Frenken, Why is economic geography not an evolutionary science? Towards an evolutionary economic geography. *J. Econ. Geogr.* **6**, 273–302 (2006).

34. R. Boschma, F. Koen, "Evolutionary economic geography" in *The New Oxford Handbook of Economic Geography*, (2018), pp. 213–229.

35. C. A. Hidalgo, R. Hausmann, The building blocks of economic complexity. *Proc. Natl.*

*Acad. Sci.* **106**, 10570–10575 (2009).

36. K. W. Boyack, R. Klavans, Creation of a highly detailed, dynamic, global model and map of science. *J. Assoc. Inf. Sci. Technol.* **65**, 670–685 (2014).

37. K. W. Boyack, R. Klavans, Co-citation analysis, bibliographic coupling, and direct citation: Which citation approach represents the research front most accurately? *J. Am. Soc. Inf. Sci. Technol.* **61**, 2389–2404 (2010).

38. K. W. Boyack, R. Klavans, K. Börner, Mapping the backbone of science. *Scientometrics* **64**, 351–374 (2005).

39. K. W. Boyack, K. Börner, R. KLAVANSb, Mapping the structure and evolution of chemistry research. 16 (2009).

40. D. A. King, The scientific impact of nations. *Nature* **430**, 311–316 (2004).

41. F. Bray, *et al.*, Global cancer statistics 2018: GLOBOCAN estimates of incidence and mortality worldwide for 36 cancers in 185 countries. *CA. Cancer J. Clin.* **68**, 394–424 (2018).

42. J. A. Evans, J.-M. Shim, J. P. A. Ioannidis, Attention to Local Health Burden and the Global Disparity of Health Research. *PLoS ONE* **9**, e90147 (2014).

43. World Health Organization, "World health statistics 2018: monitoring health for the SDGs" (2018) (August 30, 2020).

44. M. A. Serrano, M. Boguna, A. Vespignani, Extracting the multiscale backbone of complex weighted networks. *Proc. Natl. Acad. Sci.* **106**, 6483–6488 (2009).

45. V. A. Traag, L. Waltman, N. J. van Eck, From Louvain to Leiden: guaranteeing well-connected communities. *Sci. Rep.* **9**, 5233 (2019).

46. R. Klavans, K. W. Boyack, The Research Focus of Nations: Economic vs. Altruistic Motivations. *PLOS ONE* **12**, e0169383 (2017).

47. N. Fantom, U. Serajuddin, *The World Bank's Classification of Countries by Income* (The World Bank, 2016) (August 30, 2020).

48. F. L. Pinheiro, D. Hartmann, R. Boschma, C. A. Hidalgo, The time and frequency of unrelated diversification. *Res. Policy*, 104323 (2021).

49. S. Bustos, C. Gomez, R. Hausmann, C. A. Hidalgo, The Dynamics of Nestedness Predicts the Evolution of Industrial Ecosystems. *PLoS ONE* **7**, e49393 (2012).

50. J. Gao, Y.-C. Zhang, T. Zhou, Computational Socioeconomics. *Phys. Rep.* **817**, 1–104

(2019).

51. G. Morrison, *et al.*, On Economic Complexity and the Fitness of Nations. *Sci. Rep.* **7**, 15332 (2017).

52. R. Inglesi-Lotz, M. Balcilar, R. Gupta, Time-varying causality between research output and economic growth in US. *Scientometrics* **100**, 203–216 (2014).

53. P. Vinkler, Correlation between the structure of scientific research, scientometric indicators and GDP in EU and non-EU countries. *Scientometrics* **74**, 237–254 (2008).

54. A. Gelman, H. Stern, The Difference Between "Significant" and "Not Significant" is not Itself Statistically Significant. *Am. Stat.* **60**, 328–331 (2006).

55. S. M. S. Krammer, Science, technology, and innovation for economic competitiveness: The role of smart specialization in less-developed countries. *Technol. Forecast. Soc. Change* **123**, 95–107 (2017).

56. W. W. Powell, K. Snellman, The Knowledge Economy. *Annu. Rev. Sociol.* **30**, 199–220 (2004).

57. W. Glänzel, National characteristics in international scientific co-authorship relations. 47 (2001).

58. M. D. Gordin, *Scientific Babel: How science was done before and after global English* (University of Chicago Press, 2015).

59. A. Éric, V.-G. Étienne, C. Grégoire, L. Vincent, Y. Gingrasb, Benchmarking scientific output in the social sciences and humanities: The limits of existing databases. *Scientometrics* **68**, 329–342 (2006).

60. C. R. Sugimoto, L. Vincent, *Measuring research: What everyone needs to know* (Oxford University Press, 2018).

61. World Bank, World Development Indicators. (2019). GDP (current US$). https://data.worldbank.org/indicator/NY.GDP.MKTP.CD?locations=1W.

62. M. J. Barber, Modularity and community detection in bipartite networks. *Phys. Rev. E* **76**, 066102 (2007).

63. M. Almeida-Neto, P. Guimarães, P. R. Guimarães, R. D. Loyola, W. Ulrich, A consistent metric for nestedness analysis in ecological systems: reconciling concept and measurement. *Oikos* **117**, 1227–1239 (2008).